\documentclass[aps,pra,twocolumn,amsmath,amssymb]{revtex4} 
\usepackage{graphicx}
\begin{document}  

\title{Creation of discrete solitons and observation of the Peierls-Nabarro barrier in 
Bose-Einstein Condensates.}

\author{V. Ahufinger$^{1}$, A. Sanpera$^{1}$, P. Pedri$^{1,2}$, 
L. Santos$^{1}$, M. Lewenstein$^{1}$}  
\address{(1) Institut f\"ur Theoretische Physik, Universit\"at Hannover, D-30167 Hannover,Germany}
\address{(2) Dipartimento di Fisica, Universit\`a di Trento and BEC-INFM, I-38050 Povo, Italy}

\begin{abstract}  
We analyze the generation and mobility  
of discrete solitons in Bose-Einstein condensates 
confined in an optical lattice under realistic experimental conditions. 
We discuss first the creation of 1D discrete solitons, for 
both attractive and repulsive interatomic interactions. We then address
the issue of their mobility, focusing our attention on 
the conditions for the experimental observability 
of the Peierls-Nabarro barrier. Finally we report on 
the generation of self-trapped structures in two and three dimensions. 
Discrete solitons may open alternative routes for the manipulation 
and transport of Bose-Einstein condensates.
\end{abstract}  
\maketitle  


\section{Introduction}
\label{sec:1}


The experimental achievement of the Bose-Einstein condensate (BEC) \cite{BEC} 
has outbursted an extraordinary interest within the last years in the physics 
of ultracold atomic gases. This interest can be partially explained 
by the inherently nonlinear character of the BEC physics induced by 
the interatomic interactions. At sufficiently low temperatures, the 
physics of the condensates is governed by a nonlinear Sch\"{o}dinger 
equation with cubic nonlinearity, also called 
Gross-Pitaevskii equation (GPE), similar as that encountered 
in other physical systems, as e.g. 
nonlinear optics (NLO) in Kerr media. The analysis of the 
resemblances between BEC physics and NLO has lead to 
the rapidly-developing field of Nonlinear Atom Optics (NLAO) \cite{Meystrebook}. 
Recently several experiments have highlighted various NLAO phenomena, 
as dark solitons in BEC with repulsive interatomic interactions 
\cite{darkSolitons1,darkSolitons2}, bright solitons in 1D BECs 
with attractive interactions \cite{brightSolitons1,brightSolitons2}, and condensate collapse 
\cite{JILACollapse}.

During the last few years, the possibility of loading a BEC in an optical lattice 
formed by a laser standing wave has attracted considerable attention, mostly motivated  
by the close resemblance between these systems and solid-state devices. In this sense, 
several remarkable experiments have been recently reported, as the observation of Bloch oscillations 
of BECs \cite{kasevich,PisaBloch}, the realization of   
Josephson-junction arrays of BECs placed in different lattice sites \cite{JosephsonFlorence}, or 
even the achievement of the superfluid to Mott-Insulator transition \cite{Greiner}.
Recently several nonlinear BEC phenomena have been analyzed in the presence of optical lattices, 
as the dynamical superfluid to insulator transition \cite{SupIns}, the BEC transport in the presence of
dispersion managing \cite{Oberthaler1}, and the generation of gap solitons, i.e. bright solitons 
with condensates with repulsive interactions \cite{Meystre,Oberthaler2}.

Particular interest has been recently devoted to those phenomena occurring 
when the condensate dimensions become comparable to the 
lattice wavelength. In that case, the discrete structure introduced by the lattice potential may lead 
to similar phenomena as those observed in NLO in periodic structures. In particular the analysis of  
discrete solitons (DS) \cite{Christo,Eisenberg,Christo2} in the BEC context has
recently attracted a growing attention \cite{Smerzi,Abdullaev,Kivshar,Kivshar2,Salerno}. 
Specially interesting phenomena exclusively induced by the 
discreteness of the system have been analyzed in the NLO context, 
such as the restriction of the mobility of the DSs due to the 
so-called Peierls-Nabarro (PN) barrier \cite{Kivshar93}, or the possibility to generate 
two-dimensional DSs \cite{Christo3}.

Although several properties related with DSs in BEC have been already reported 
\cite{Smerzi,Abdullaev,Kivshar}, the realization of DSs under realistic experimental 
conditions has so far not been analyzed in detail. Therefore, one of the aims of the 
present paper is to discuss the creation of these structures in the frame 
of the recent experiments on BEC in optical lattices. In particular, we shall 
discuss the generation of 1D DS for both attractive and repulsive interacting condensates.
Once created, the effects of the discrete nature of the DSs in BEC should be 
analyzed by means of the observation of the PN barrier for its mobility. 
A second aim of the present paper is to discuss the conditions for the experimental 
observability of this barrier. Interestingly, the PN barrier is 
largely overestimated within the usual tight-binding approximation even under conditions for which 
this approximation is typically assumed. Finally, in the last part of our 
paper, we discuss the possibility of achieving 2D and even 3D self-trapped structures, 
which could offer alternative routes for the controllable manipulation of BECs. 

The scheme of the paper is as follows. In Sec.~\ref{sec:2} we discuss the physical system 
under consideration as well as the basic equations to describe it. 
Sec.~\ref{sec:3} presents a variational approach which allows for an analysis of 
discrete structures in arbitrary dimensions.
Sec.~\ref{sec:4} is devoted to the analysis of the generation 
of 1D DSs, for both attractive and repulsive 
interatomic interactions. We address also in Sec.~\ref{sec:4} the issue of their 
mobility and provide the conditions for the observability of the Peierls-Nabarro barrier. 
In Sec.~\ref{sec:5} we analyze the creation of 2D and 3D 
self-trapped structures. We finalize in Sec.~\ref{sec:6} with our conclusions.

\section{Physical system} 
\label{sec:2}

In the following we consider a trapped BEC in the presence of an optical lattice. 
The periodic structure leads to an energy band structure 
\cite{Sorensen,Javanainen,Choi,Chiofalo}, and strongly modifies the dynamics of the condensate 
\cite{kasevich,PisaBloch,JosephsonFlorence,SupIns,Oberthaler1,Meystre,Oberthaler2,Stringari}. 
In the mean field approximation, the full BEC dynamics (at a temperature much smaller than 
the critical one for the condensation) is governed by the time-dependent GPE:
\begin{equation}
i\hbar \frac{d\psi \left(\vec{r},t\right)}{dt} = \left(-\frac{\hbar^2}{2m}\triangle+
V \left(\vec{r}\right)+g\vert \psi\left(\vec{r},t\right)\vert^2\right)\psi \left(\vec{r},t\right),
\label{GPE}
\end{equation}
where $g=4\pi\hbar^2a/m$, with $a$ the $s$-wave scattering length and 
$m$ the atomic mass. The condensate wave function is normalized to the total number of particles $N$. 
The external potential is given by:
\begin{eqnarray}
V \left(\vec{r} \right)=\frac{m}{2} \left({\omega_x}^2 x^2+{\omega_y}^2 y^2+{\omega_z}^2 z^2\right)+
\nonumber \\
V_0 \left(\sin^2\left(\frac{\pi x}{d}\right)+\sin^2\left(\frac{\pi y}{d}\right)
+\sin^2\left(\frac{\pi z}{d}\right)\right),
\end{eqnarray}
which describes both the magnetic trap potential and the optical lattice 
(created by two counter propagating laser beams 
of wavelength $\lambda$ along each axis). 
The angular frequencies of the magnetic trap in each direction
are denoted by  $\omega_i$. The optical lattice is characterized  by its depth $V_0$ and by its lattice
period $d=\lambda/2$, which defines the so-called recoil momentum $k_r=\frac{\pi}{d}$. 
In the following, and following the standard notation, we refer 
the depth of the optical potential in units of the so-called 
recoil energy $E_r=\hbar^2 k_r^2/2m$. 

When the optical depth of the lattice 
is much larger than the chemical potential ($V_0>>\mu$), and the system can be considered 
as confined within the lowest energy band, one can employ the tight-binding approximation 
and rewrite the condensate order parameter 
as a sum of wavefunctions localized in each well of the periodic potential:
\begin{equation}
\psi \left(\vec{r},t\right)=\sqrt{N}\sum_{n} \phi_{n} \left(t\right)\varphi_{n} \left(\vec{r}\right),
\label{tigth}
\end{equation}
where $\varphi_{n} \left(\vec{r}\right)=\varphi \left(\vec{r}-\vec{r}_n\right) $ denotes the on-site wavefunction.
By inserting the Ansatz (\ref{tigth}) into Eq.(\ref{GPE}), one obtains that the GPE reduces to a 
discrete nonlinear Schr\"odinger equation (DNLSE): 
\begin{equation}
i\hbar \frac{\partial\phi_{n}}{\partial t} =-J \left(\phi_{n-1}+\phi_{n+1}\right)+\left(\epsilon_{n}+U \vert \phi_{n}\vert^2\right)\phi_{n}.
\end{equation}
The dynamics of the system depends mostly on the interplay between the tunneling rate (J) and the two body interactions (non linear energy, U). The tunneling rate can be expressed as:
\begin{equation}
J=-\int d\vec{r} \,\large\lbrack\frac{\hbar^2}{2m}\vec{\triangledown}\varphi_{n}\vec{\triangledown}\varphi_{n+1}+\varphi_{n}V\left(\vec{r} \right)\varphi_{n+1}\large\rbrack,
\label{tun}
\end{equation}
the nonlinear term acquires the form:
 \begin{equation}
U=gN\int d\vec{r}\varphi_{n}^4
\label{nolin}
\end{equation}
and the on-site energies are given by:
\begin{equation}
\epsilon_{n}=\int d\vec{r}\,\large\lbrack\frac{\hbar^2}{2m}\left(\vec{\triangledown}\varphi_{n}\right)^2+V\left(\vec{r} \right)\varphi_{n}^2\large\rbrack .
\end{equation}
In order to calculate the value of the nonlinear energy Eq.~(\ref{nolin}), 
we use a Gaussian Ansatz for the wavefunction on-site, where the width 
is obtained by minimization of the energy ~\cite{Paolo}. 
In the pure one-dimensional case, we use the interaction constant 
obtained by averaging the 3D coupling constant over the radial density profile ~\cite{Olshanii}. 
To calculate the tunneling rate, the same Gaussian Ansatz can be employed in 
Eq.~(\ref{tun}) or, in the limit $V_0>>E_r$, it can be 
obtained from the exact result for the width of the lowest band in the 1D 
Mathieu-equation ~\cite{Zwerger}.   

\section{variational calculation} 
\label{sec:3}
\begin{figure}
\includegraphics[width=1.0\linewidth]{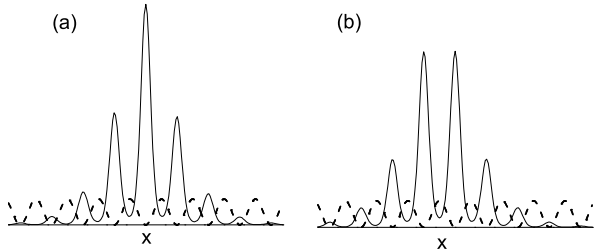}
\caption{Density profile of a discrete soliton (solid line) with (a) the center of mass centered in one minimum of the optical lattice and (b) the center of mass displaced by half lattice period.
}\label{barrera}
\end{figure}

Discrete solitons are characterized by being stable solutions of the Hamiltonian 
which propagate without distortion. They correspond to minima of the energy of the system which  
in the tight-binding approximation acquires the form: 
\begin{equation}
E=\sum_{n=-\infty}^{\infty}\lbrace -J\phi_{n}^*\left(\phi_{n-1}+\phi_{n+1}\right)+
\epsilon_n \vert \phi_{n}\vert^2+\frac{U}{2}|\phi_{n}\vert^4\rbrace .
\label{varenergie}
\end{equation}
Let us consider first the 1D case. 
To find the minima of the energy, an exponential Ansatz 
for the soliton envelope given by $\phi_{n}=C\exp{\left(-\beta\vert n \vert \right)}$ can be employed, 
where $C$ is a normalization constant and $\beta$ is a variational parameter 
which accounts for the inverse width of the soliton. 
Introducing this Ansatz in Eq.(\ref{varenergie}) the expression for the energy, 
in the negative scattering length case, becomes:
\begin{equation}
\frac{E}{\vert U \vert}=-\frac{4J}{\vert U \vert}\frac{e^\beta}{e^{2\beta}+1}-\frac{1}{2}
\frac{\left(e^{2\beta}-1\right)\left(e^{4\beta}+1\right)}{\left(e^{2\beta}+1\right)^3}. 
\label{mincentered}
\end{equation}
By minimizing $E$ with respect to the inverse width $\beta$ 
one obtains the energy of the discrete structure
centered in one minimum of the optical lattice (Fig.\ref{barrera}(a)).

Notice that in order to ensure that the discrete structure corresponding to 
this minimum is indeed a soliton, one has to address the issue 
of its mobility. In contrast with the case of continuous solitons where applying 
an external momentum results in a linear response of the soliton,  
in discrete systems a similar effect occurs only for broad soliton 
distributions (i.e., those that  occupy many sites). 
Conversely if the dimensions of the soliton are of the order of 
the lattice wavelength, $\lambda$, the discreteness of the system begins to play 
a fundamental role. In particular, the discreteness generates an effective 
periodic potential energy, whose amplitude is the minimum barrier which must be overcome 
to translate the center of mass of the system half a lattice period, i.e. 
from one minimum of the lattice (Fig.\ref{barrera}(a)) 
to a neighboring lattice maximum (Fig.\ref{barrera}(b)). This is the previously mentioned 
PN barrier \cite{Kivshar93}.  
The energy of the state  depicted in Fig.\ref{barrera}(b) can be also calculated within a 
variational approach 
using $\phi_{n}=C\exp{\left(-\beta\vert n-\frac{1}{2} \vert \right)}$ as an Ansatz:
\begin{equation}
\frac{E}{\vert U \vert}=-\frac{2J}{\vert U \vert}\frac{\left(1+\sinh\left(\beta\right)\right)}{e^{\beta}}-\frac{1}{4}\frac{\left(e^{2\beta}-1\right)}{\left(e^{2\beta}+1\right)}.
\label{minnoncentered}
\end{equation}
Again by minimization one obtains the energy of the displaced discrete structure.
The difference between both energies 
(minima of Eq.(\ref{mincentered}) and Eq.(\ref{minnoncentered})) corresponds to the PN barrier. 

Notice that the barrier becomes relevant when the soliton structure occupies
few sites of the lattice. By increasing the number of occupied states the above 
two modes approach in energy and the barrier decreases approaching zero as the number of sites grows.
In this sense, the PN barrier is a distinctive discrete phenomena. 

The above analysis can be straightforwardly generalized to higher dimensions. 
Assuming the most general case, in which the width of the exponential Ansatz 
along the direction of the movement 
is different from the other directions, it can be shown that localized structures are also possible in two 
and three dimensions under an appropriate ratio between tunneling and nonlinear energy 
(see Sec.~\ref{sec:5}). 
A variational calculation using a Gaussian Ansatz instead of the exponential one has also been performed 
obtaining equivalent results.

\section{Generation and movement of one-dimensional discrete solitons} 
\label{sec:4}

\begin{figure}
\includegraphics[width=0.8\linewidth]{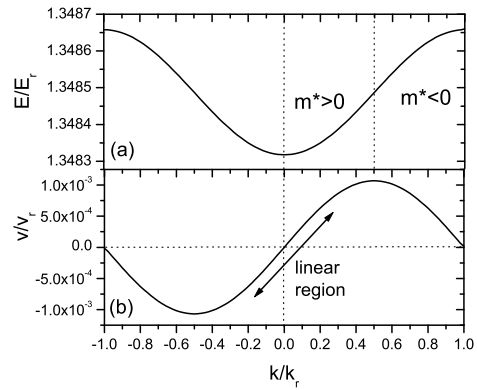}
\caption{(a)Energy of the first band in units of the recoil energy and (b)velocity profile in units of $v_r=\frac{\hbar \pi}{m d}$ as a function of the quasimomentum in the first Brillouin zone for an optical potential depth of $8E_r$}
\label{banda}
\end{figure}

In this section we analyze the issue of the generation and mobility of DSs in 1D BEC under realistic 
experimental conditions. Let us first discuss the generation of DSs 
in condensates with positive scattering length. In particular, we 
consider a $^{87}$Rb condensate hold in a magnetic trap and 
in the presence of an optical lattice. As discussed in \cite{Smerzi,Abdullaev} it is possible  
to generate DSs in pure 1D repulsive condensates 
when the tunneling rate balances the nonlinear energy of the system.  
For positive scattering lengths, the compensation of these two effects 
is not possible unless the system has a negative effective mass $m^*$
($\frac{1}{m*}=\frac{1}{\hbar^2} \frac{\partial^2 E}{\partial k^2}$ where $E$ is the energy of the first band and $k$ its quasimomentum (see Fig.~\ref{banda})). 
At least two possible mechanisms can easily place the system in such a 
region, inverting thus the sign of the tunneling. One possibility
\cite{Oberthaler1} consists on providing the condensate with an external 
momentum to place it at the edge of the first Brillouin zone 
where the effective mass is negative (see Fig.~\ref{banda}(a)). This can be done 
for instance, by introducing  --in the absence of the magnetic trap --
a tilt in the optical lattice.
A second mechanism to reach the negative effective mass region relies 
on the variation of the relative phase of the condensate in the lattice. 
Concretely, a repulsive condensate in a periodic potential with a phase 
difference of $\pi$ between consecutive wells gives rise to 
the so-called staggered discrete soliton type \cite{staggered}. 
Such a phase structure can be achieved using the well established 
method of phase imprinting \cite{dobrek} which allows to modify the phase 
of a condensate without modifying its density profile.
To this aim we propose to use a second optical lattice with double spatial period 
than the first one acting for a time much
shorter than the characteristic times of the system, i.e. the 
correlation ($\tau_c=\frac{\hbar}{\mu}$) 
and the tunneling ($\tau_t=\frac{\hbar}{J}$) time. 
The phase imprinted in this way depends solely on the amplitude of this 
second standing wave and on the time in which it acts.
\begin{figure}
\includegraphics[width=0.8\linewidth]{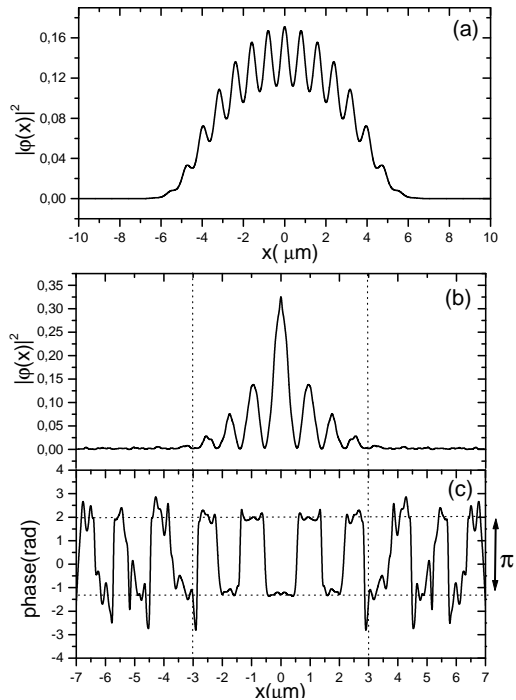}
\caption{(a) Density profile of the ground state of a ${^{87}}$Rb condensate in the combined (magnetic and optical) trap for $ N=2000$, $\omega_x=50\times 2\pi$ Hz, $\omega_y=\omega_z=92\times 2\pi$ Hz, $a=5.8$nm, $V_0=1E_r$ and $d=0.8\mu$m.(b) Density profile of the staggered soliton 100ms after the imprinting of a phase difference of $\pi$ between consecutive wells and after the magnetic trap is switched off.(c) Phase profile of the staggered soliton shown in (b)}
\label{rubidi}
\end{figure}
In the following we describe in detail how the phase imprinting 
method is implemented. Firstly we calculate the ground state of the 
system in the presence of an optical lattice. 
This can be done either by calculating directly 
(using GPE in imaginary time) the
ground state of the system in the presence of both the magnetic trap and
the optical lattice or by calculating the ground state of the system in the
presence of the magnetic trap only and afterwards growing adiabatically 
the optical lattice and letting the system to evolve to the new ground state.
As expected both methods yield the same ground state.
Once the ground state is found, a second optical lattice 
with an amplitude of  $72.5E_r$ acting for $t=0.4$ms performs the phase 
imprinting while the magnetic trap is suddenly switched off.  
Our results are summarized in Fig.~\ref{rubidi}. Fig.~\ref{rubidi}(a)  
shows the density profile of the ground state of the combined trap (magnetic and lattice) while
Fig.~\ref{rubidi}(b) displays the density profile of the localized structure $100$ms 
after turning off the magnetic trap. 
The system evolves from the ground state shown in 
Fig.~\ref{rubidi}(a) to the localized structure shown in Fig.~\ref{rubidi}(b) 
which remains unaltered for times much longer 
than the tunneling time of the system. This structure contains 40\% of the initial number of atoms. In Fig ~\ref{rubidi}(c), 
we display the phase profile corresponding to 
Fig.~\ref{rubidi}(b) in where clear phase jumps of $\pi$ between consecutive sites of the optical lattice 
in the spatial region occupied by the localized structure are present.
Finally by applying an external momentum to the localized structure we observe that it moves without distortion, evidencing thus that such structures correspond indeed to discrete bright solitons. 
Notice that the applied momentum must keep the structure in the negative effective mass 
region of the band (Fig.~\ref{banda}(a)) in order not to destroy the soliton. 

We turn now to the case of negative scattering length. 
In continuous systems, the attractive nature of the interactions compensates 
the effect of the kinetic energy and, therefore,  
the ground state of a one dimensional homogeneous condensate 
with negative scattering length is already a bright soliton. 
The presence of an optical lattice in such systems 
permits the creation of a discrete bright soliton. 
Thus, for a fixed number of atoms, $N$, one can 
straightforwardly vary the number of occupied sites by varying only 
 the depth of the optical potential, $V_0$.
\begin{figure}
\includegraphics[width=1.0\linewidth]{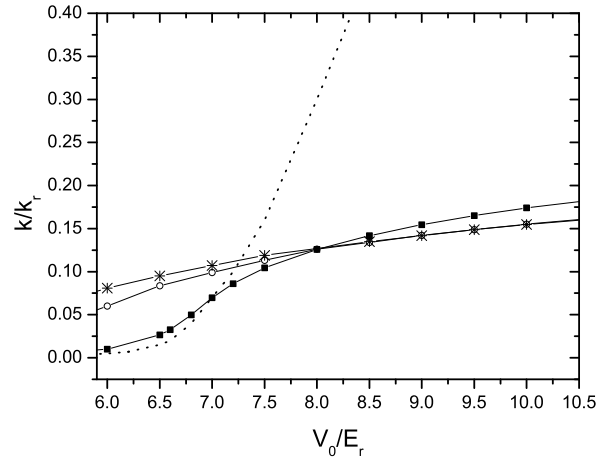}
\caption{Momentum associated with the PN
 barrier calculated with the variational method (open dots), with the imaginary 
time evolution of the GPE (squares) and with the DNLSE (stars). The necessary momentum to move the soliton (calculated with the real time evolution of the GPE) is displayed by the dotted line.}\label{PN}
\end{figure}
Since the interactions are now attractive 
the number of sites occupied by the discrete solitons 
is typically much smaller than in the case of positive 
scattering length. Bearing all this features in mind, it becomes clear 
that such systems are unique to study how the movement of a 
discrete soliton depends on the degree of discreteness of the structure, i.e.,   
to observe the PN barrier ~\cite{Kivshar93}. 
We have analyzed the case of a condensate of $^7$Li confined in a magnetic trap with 
$\omega_y=\omega_z=20\times 2\pi$ kHz, and $\omega_x=75\times 2\pi$ Hz. 
We fix $aN=-14$nm~\cite{nota}. 
Again, to analyze the generation and the propagation of 
discrete structures we start by calculating  the ground state of the system 
in the combined (lattice + magnetic trap) potential.
We proceed as in the case of Rubidium, either growing adiabatically 
the optical lattice after the ground state of the condensate in the trap 
has been found or by finding directly the ground state in the presence 
of both potentials.  After the magnetic trap is switch off we observe
that the localized structure remains without dispersion for times much 
larger than the tunneling time of the system. 
To study the effect of the discreteness we vary the values of the depth of
the optical potential between $6\leq {V_0}/{E_r}\leq 10$ for a fixed lattice spacing of $d=0.8\mu $m. 
The number of sites significantly occupied by the discrete structure vary then between 11 and 3.

We calculate for these sets of parameters the PN 
barrier, i.e., the difference in energy between the localized structure
centered in a minimum of the periodic potential (Fig.~\ref{barrera}(a)) and the structure 
corresponding to a lattice displaced by half lattice period (Fig.~\ref{barrera}(b)).
To test the validity of the variational method and the 
validity of the tight binding approximation, we calculate the PN barrier
using the expressions developed in Sec.~\ref{sec:3} and solving numerically in 
imaginary time the DNLSE (using a Runge-Kutta method). Then, we compare these results 
with the barrier obtained by solving directly the full GPE in imaginary time.
The results of those calculations are summarized in Fig.~\ref{PN}, where we display the
momentum associated with the PN barrier (in units of $k_r$) as a function of the depth $V_0$ of the
optical lattice.

We observe that for large optical potential depths 
($V_0\geq 8E_r$) the results obtained with the DNLSE and the variational method
using the exponential Ansatz match each other perfectly and agree well with 
those obtained by solving directly the GPE. For low optical potential depths 
($V_0<8E_r$) the DNLSE and the variational start to disagree because the exponential 
Ansatz becomes less appropriate as the number of sites increases. 
Moreover, in this region a clear disagreement between the GPE and the methods 
that assume tight binding appears, evidencing thus the inapplicability of the 
tight binding approximation. Note that, surprisingly, the tight-binding results 
for the PN barrier fail to match with the GPE solution even for values of the lattice potential that are 
typically assumed as being well into the tight-binding regime. In this sense 
our calculations reveal that the PN barrier is certainly very sensitive 
to slight deviations from the tight-binding conditions. The latter can be 
explained taking into account that  
if the on-site dynamics inside each potential well is not completely frozen, as assumed 
in Eq.~(\ref{tigth}), the energy associated to the on-site movement 
can smear out the PN barrier.

\begin{figure}
\includegraphics[width=0.8\linewidth]{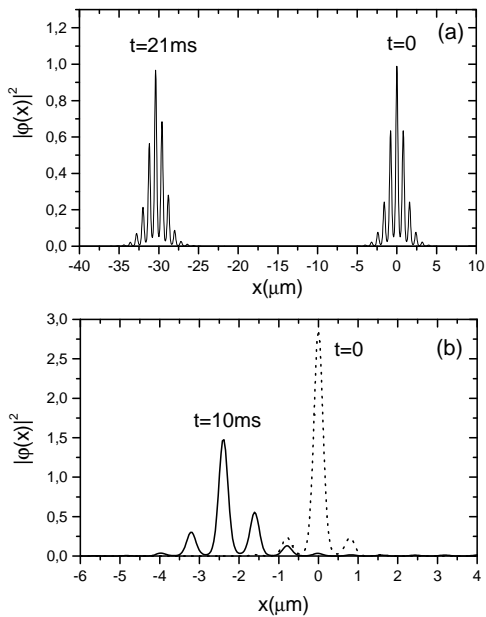}
\caption{(a)Density profile of a discrete soliton created in the presence of an optical potential of depth $6E_r$ before and 21ms after a transfer of momentum of $0.1k_r$ is applied to the structure. (b) Density profile of the localized structure generated with a lattice of amplitude $8E_r$ (dotted line) and the corresponding density profile 10ms after a kick of $0.3k_r$ is given (solid line). $77$$\%$ of the atoms remain in the reconfigurated structure.}
\label{mov}
\end{figure}

Let us recall that physically the presence 
of the PN potential implies that the soliton will only move if a momentum above
the critical one, determined by the PN barrier, is provided. 
Therefore, we perform a dynamical study by solving the GPE in real time 
after providing an instantaneous transfer of momentum to the structure. The latter 
can be achieved either via phase imprinting or by applying a linear potential 
during a time shorter than any other time scale. We monitor for 
every value of $V_0$ the minimum applied momentum to move the localized structure.
The results are displayed in Fig.~\ref{PN} as a dotted line. 
We observe that the barrier calculated with the static GPE and the dynamical results only agree for 
low optical potential depths ($V_0<7E_r$). For such cases 
the soliton is spread considerably in real space being, therefore, 
well localized in momentum space. 
This is the necessary condition to move the structure as a whole 
in the linear region of the velocity profile (Fig.~\ref{banda}(a)) 
where the velocity is proportional to the given momentum. The movement for the 
case of $V_0=6E_r$ after a transfer of momentum of $0.1k_r$ is depicted in 
Fig.~\ref{mov}(a). Conversely, a clear deviation from all the previous calculations 
of the PN barrier appears as the lattice potential depth increases.
Our results show that there exist regions in which it is not possible
to reach dynamically the configuration corresponding to Fig.~\ref{barrera}(b). 
This is what happens for large optical 
potentials ($V_0>8E_r$) where the soliton is highly localized in space 
and, therefore, spread in momentum. We observe that for ($V_0\geq10E_r$), 
no motion is found no matter how large is the initial momentum given to
the soliton. In this case, the width in momentum space 
of the localized structure is of the order of the momentum of the lattice, 
i.e., the first Brillouin zone is saturated and 
the movement is prevented. 
For intermediate optical potentials ($8-9E_r$), we observe 
that movement of these structures is not possible in the linear part 
of the velocity profile (Fig.~\ref{banda}) even if one overcomes the PN barrier. 
Nevertheless, if the given momentum is large enough, 
we observe in our simulations a reorganization of the structure which allows it to move 
by losing first a fraction of the atoms and spreading afterwards in space, being later on enough localized 
in momentum space to move as a whole (Fig.~\ref{mov}(b)). The re-arranged structure 
exhibits a complex dynamics as a result of the interplay between nonlinearity and band structure. 

\section{Creation of two- and three-dimensional self-trapped structures} 
\label{sec:5}
\begin{figure}
\includegraphics[width=1.0\linewidth,clip]{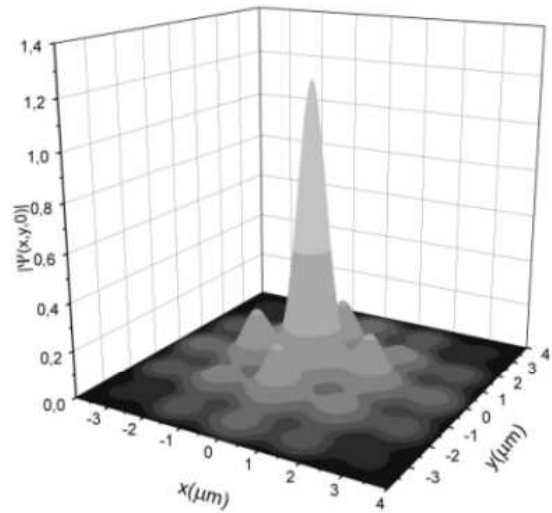}
\caption{Spatial structure of a matter discrete localized structure created in the presence of 
a 3D optical lattice with a depth $ V_0=6E_r$, and a period $d=1.6\mu$m, 
as a function of two spatial coordenates, (x,y), for a fixed value of the third coordinate, z=0.}
\label{highdim}
\end{figure}
In this section, we briefly address the experimental possibilities for the generation of 
discrete solitons in higher spatial dimensions in attractive condensates \cite{Kivshar2,Salerno}. 
From the variational Ansatz discussed in 
Sec.~\ref{sec:3}, it is possible to obtain that two, and three-dimensional 
self-trapped structures  may be stable provided that the ratio $|J/U|$ is low enough. 
The threshold value for the 2D case is $0.175$($0.2$) and for the $3D$ case 
$0.13$($0.15$) using an exponential (Gaussian) Ansatz. The limitation to such low tunneling 
rates imposes a restriction in the maximal number of sites occupied by the localized structure. 

We consider the full 3D case with an spherical magnetic trap and we address 
the creation of 3D localized structures in the presence of a 2D or 3D optical lattice. 
Notice that due to the attractive interactions, the possibility of collapse imposes 
some restrictions in the number of atoms and in the features of the optical lattice. 
In some situations, even if the collapse in the magnetic trap is prevented, 
the adiabatic growing of the lattice leads to the on-site collapse. This effect must be 
taken into account if the 3D localized structures are to be created. 
A condensate of $^7$Li with $aN=-70$nm and 
initially in a spherical magnetic trap of frequencies 
$\omega_x= \omega_y=\omega_z=375\times 2\pi$Hz has been considered. 
This situation corresponds for instance to $N=500$ $^{7}Li$ atoms and a 
modified $s$-wave scattering length $a=-0.14$nm available with 
Feshbach resonances ~\cite{brightSolitons1,brightSolitons2}.
Fig.~\ref{highdim} displays the spatial structure of a matter 
discrete localized structure created in the presence of a 
3D optical lattice with period $d=1.6\mu$m, and $ V_0=6E_r$. 

Therefore, the discreteness of the lattice potential allows for the interesting 
possibility of a controllable generation of a self-trapped regular 3D BEC structure. 
Unfortunately, the restriction in the number of sites occupied in the localized structures 
created in the two and three dimensional cases, imposes a severe limitation 
for their mobility. As already discussed for the 1D case, 
the spread in momentum space of these structure prevents their motion, 
at least in the linear region of the band structure. 

\section{Conclusions} 
\label{sec:6}
We have analyzed the conditions to generate discrete solitons in 1D Bose-Einstein condensates, 
either with positive or negative scattering length. In particular, in 
repulsive interacting condensates, the phase imprinting method 
has been proposed to controllably create bright staggered type solitons. 
Once generated, we have addressed the mobility of these structures. This mobility is 
characterized by two different effects: (i) the presence of the PN barrier, 
a purely discrete effect which sets a minimal kinetic energy to move 
half a lattice period, and (ii) the spreading of the atomic wavefunction in momentum 
space due to the spatial localization in few lattice sites. The mobility is only possible 
if the PN barrier is overcome, but even if this is the case a clear soliton 
motion is only possible if the system is placed in a region of linear dispersion. 
Our analysis shows that the estimation of the PN barrier is crucially sensitive 
to slight deviations from the tight-binding conditions. In particular, the 
tight-binding approximation has been shown to fail significantly for large lattice 
potentials which are typically assumed to guarantee the validity of such an approximation.

The mobility of discrete solitons generated in this way offer interesting possibilities in the 
context of BEC guiding. In this sense similar ideas as those analyzed in the 
case of optical DS \cite{Christoguides} could be employed to generate DS networks, which 
could consitute a novel (non dispersive) approach to the issue of integrated atom optics. 

Finally, we have shown that the discreteness of the lattice also allows for self-trapped 
structures in 2D and 3D. Due to the strong localization required, 
their mobility is prevented to a large extent. 
These higher-dimensional structures could open alternative routes 
to optical tweezers \cite{Ketterle} and magnetic conveyor belts \cite{Conveyor},  
for the storing, manipulation and transport of atomic condensates.

We acknowledge support from Deutsche Forschungsgemeinschaft (SFB 407),  
the RTN Cold Quantum gases, ESF PESC BEC2000+, and the Ministero dell'Istruzione, 
dell'Universit\`a e della Ricerca (MIUR).
V.A. acknowledges support from the European Community through a Marie Curie Fellowship 
(HPMF-CT-2002-01847). L. S. and P. P. wish to thank the Alexander von Humboldt Foundation, 
the Federal Ministry of Education and 
Research and the ZIP Programme of the German Government.


\begin{references}

\bibitem{BEC}M.~H.~Anderson {\it et al.}, Science {\bf 269}, 198 (1995); K.~B.~Davis {\it et al.}, Phys. Rev. Lett. {\bf 75}, 3969 (1995); C.~C.~Bradley {\it et al.}, Phys. Rev. Lett. {\bf 75}, 1685 (1995); C.~C.~Bradley,C.~A.~Sackett, R.~G.~Hulet, Phys. Rev. Lett. {\bf 78}, 985 (1997).
\bibitem{Meystrebook}P. Meystre, {\it Atom Optics}, Springer Verlag, New York, 2001.
\bibitem{darkSolitons1}S.~Burger {\it et al.}, Phys. Rev. Lett. {\bf 83}, 5198 (1999).
\bibitem{darkSolitons2}J.~Denschlag {\it et al.}, Science {\bf 287}, 97 (2000).
\bibitem{brightSolitons1}L.~Khaykovich {\it et al.}, Science {\bf 296}, 1290 (2002).
\bibitem{brightSolitons2}K.~E.~Strecker {\it et al.}, Nature {\bf 417}, 150 (2002).
\bibitem{JILACollapse}E.~A.~Donley {\it et al.}, Nature {\bf 412}, 295 (2001).
\bibitem{kasevich}B.~P.~Anderson and M.~A.~Kasevich, Science {\bf 282}, 1686 (1998). 
\bibitem{PisaBloch}O.~Morsh {\it et al.}, Phys. Rev. Lett. {\bf 87}, 140402 (2001).
\bibitem{JosephsonFlorence}F.~S.~Cataliotti {\it et al.}, Science {\bf 293}, 843 (2001).
\bibitem{Greiner} M. Greiner {\it et al.}, Nature (London) {\bf 415}, 39 (2002).
\bibitem{SupIns} P. G. Kevrekidis {\it et al.}, Phys. Rev. Lett. {\bf 89}, 170402 (2002); 
F. S. Cataliotti {\it et al.}, New J. Phys. {\bf 5}, 71 (2003). 
\bibitem{Oberthaler1} B.~Eiermann {\it et al.}, Phys. Rev. Lett. {\bf 91}, 060402 (2003).
\bibitem{Meystre}O.~Zobay {\it et al.}, Phys. Rev. A {\bf 59}, 643 (1999).
\bibitem{Oberthaler2} Gap solitons have been recently observed at the University of Konstanz, 
M. Oberthaler, private communication.
\bibitem{Christo}D.~N.~Christodoulides and R.~I.~Joseph, Opt. Lett. {\bf 13}, 794 (1988).
\bibitem{Eisenberg}H.~S.~Eisenberg {\it et al.}, Phys. Rev. Lett. {\bf 81}, 3383 (1998).
\bibitem{Christo2}J.~W.~Fleischer {\it et al.}, Phys. Rev. Lett. {\bf 90}, 023902 (2003).
\bibitem{Smerzi}A.~Trombettoni and A.~Smerzi, Phys. Rev. Lett. {\bf 86}, 2353 (2001).
\bibitem{Abdullaev}F.~Kh.~Abdullaev {\it et al.}, Phys. Rev. A {\bf 64}, 043606 (2001).
\bibitem{Kivshar} P. J. Y. Louis {\it et al.}, Phys. Rev. A {\bf 67}, 013602 (2003).
\bibitem{Kivshar2} E. ~A.~Ostrovskaya and Y.~S.~Kivshar, Phys. Rev. Lett. {\bf 90}, 160407 (2003).
\bibitem{Salerno} B. B. Baizakov, B. A. Malomed, and M. Salerno, Europhys. Lett., 63, 642 (2003).
\bibitem{Kivshar93} See Y.~S.~Kivshar and D.~K.~Campbell, Phys. Rev. E {\bf 48}, 3077 (1993) 
and references therein.
\bibitem{Christo3}J.~W.~Fleischer {\it et al.}, Nature {\bf 422}, 147 (2003).
\bibitem{Sorensen}K.~Berg-Sorensen and K.~M\o lmer, Phys. Rev. A {\bf 58}, 1480 (1999).
\bibitem{Javanainen}J.~Javanainen, Phys. Rev. A {\bf 60}, 4902 (1999).
\bibitem{Choi}D.~Choi and Q.~Niu, Phys. Rev. Lett. {\bf 82}, 2022 (1999).
\bibitem{Chiofalo}M.~L.~Chiofalo, S.~Succi and M.~P.~Tosi, Phys. Rev. A {\bf 63}, 063613 (2001).
\bibitem{Stringari} M. Kr\"amer, L. Pitaevskii, and S. Stringari, Phys. Rev. Lett. 
{\bf 88}, 180404 (2002).
\bibitem{Paolo} P. Pedri {\it et al.}, Phys. Rev. Lett. {\bf 87}, 220401 (2001).
\bibitem{Olshanii}M.~Olshanii, Phys. Rev. Lett. {\bf 81}, 938 (1998). 
\bibitem{Zwerger}W.~Zwerger, J. Opt. B: Quantum Semiclass. Opt. {\bf 5}, 9 (2003).
\bibitem{staggered} Y.~S.~Kivshar, Opt. Lett. {\bf 18}, 1147 (1993).
\bibitem{dobrek}\L.~Dobrek {\it et al.}, Phys. Rev. A {\bf 60}, R3381 (1999).
\bibitem{nota}Notice that in the simulations the fixed parameter is $Na$ and therefore, a certain freedom in the choice of the number of atoms and s-wave scattering length exists. Note that the scattering properties of $^7$Li can be strongly modified by the use of Feshbach resonances \cite{brightSolitons1,brightSolitons2}. 
\bibitem{Christoguides} D. N. Christodoulides and E. D. Eugenieva, 
Phys. Rev. Lett. {\bf 87}, 233901 (2001).
\bibitem{Ketterle} T. L. Gustavson {\it et al.}, Phys. Rev. Lett. {\bf 88}, 020401 (2002). 
\bibitem{Conveyor} M. Greiner {\it et al.}, Phys. Rev. A {\bf 63}, 031401 (2001). 


\end{references}
\end{document}